\documentclass[aps,prl,showpacs,twocolumn,superscriptaddress,nofootinbib,floatfix,10pt]{revtex4}
\usepackage{amsfonts,amssymb,stmaryrd,latexsym,amsmath}
\usepackage{graphicx,subfigure}
\usepackage{times}
\usepackage{slashed}

\begin{document}

\title{\textit{Ab initio} calculation of the spectrum and structure of $^{16}$O}

\author{Evgeny Epelbaum}\affiliation{Institut~f\"{u}r~Theoretische~Physik~II,~Ruhr-Universit\"{a}t~Bochum,
D-44870~Bochum,~Germany}

\author{Hermann~Krebs}
\affiliation{Institut~f\"{u}r~Theoretische~Physik~II,~Ruhr-Universit\"{a}t~Bochum,
D-44870~Bochum,~Germany}

\author{Timo~A.~L\"{a}hde}
\affiliation{Institute~for~Advanced~Simulation, Institut~f\"{u}r~Kernphysik, and
J\"{u}lich~Center~for~Hadron~Physics,~Forschungszentrum~J\"{u}lich,
D-52425~J\"{u}lich, Germany}

\author{Dean~Lee}
\affiliation{Department~of~Physics, North~Carolina~State~University, Raleigh, 
NC~27695, USA}

\author{Ulf-G.~Mei{\ss }ner}
\affiliation{Institute~for~Advanced~Simulation, Institut~f\"{u}r~Kernphysik, and
J\"{u}lich~Center~for~Hadron~Physics,~Forschungszentrum~J\"{u}lich,
D-52425~J\"{u}lich, Germany}
\affiliation{Helmholtz-Institut f\"ur Strahlen- und
             Kernphysik and Bethe Center for Theoretical Physics, \\
             Universit\"at Bonn,  D-53115 Bonn, Germany}
\affiliation{JARA~-~High~Performance~Computing, Forschungszentrum~J\"{u}lich, 
D-52425 J\"{u}lich,~Germany}

\author{Gautam~Rupak}
\affiliation{Department~of~Physics~and~Astronomy, Mississippi~State~University, Mississippi State, MS~39762, USA}

\begin{abstract}
\noindent
We present \textit{ab initio} lattice calculations of the low-energy even-parity states of $^{16}$O using chiral nuclear effective field theory. We find 
good agreement with the empirical energy spectrum, and with the electromagnetic properties and transition rates. For the ground state, we find that the
nucleons are arranged in a tetrahedral configuration of alpha clusters. For the first excited spin-0 state, we find that the predominant structure is a 
square configuration of alpha clusters, with rotational excitations that include the first spin-2 state.        
\end{abstract}


\maketitle  
The most abundant nucleus by weight in the Earth's crust is $^{16}$O, which also forms a key ingredient of life as we know it.
In addition to its ubiquity and central role as a life-generating element, the spectrum and structure of $^{16}$O presents several 
long-standing puzzles in nuclear physics. In the nuclear shell model, the ground state of $^{16}$O with spin-parity $J^p = 0^+$ consists 
of doubly-closed $p$-shells. Recently, several \textit{ab initio} calculations have improved on the shell-model description of the ground state of
$^{16}$O~\cite{Hagen:2010gd,Roth:2011ar,Hergert:2013uja}. Still, a number of key features of the $^{16}$O spectrum remain difficult
to address within a shell-model description. One such difficulty is that the first excited state has $0^+$ spin-parity 
quantum numbers~\cite{Nagai:1962}. Another puzzling feature is presented by the pattern of higher-spin excitations which include the
lowest spin-2 state, indicating possible rotational bands of deformed states. 

Since the early work of Wheeler~\cite{Wheeler:1937zza}, there have been theoretical studies of $^{16}$O based on alpha cluster 
models~\cite{Dennison:1954zz,Iachello,Robson:1979zz,Bauhoff:1984zza,Filikhin:1999,Tohsaki:2001an,Bijker} and some 
experimental evidence for alpha-particle states in $^{16}$O from the analysis of decay products~\cite{Freer:2005ia}.  
The case for an alpha-$^{12}$C resonant cluster structure of the excited rotational band in $^{16}$O was established in Ref.~\cite{Buck:1975zz}.  While such models have been able to describe some of the puzzles in the structure of $^{16}$O on a phenomenological (or geometrical) level, there has so far 
been no support for the alpha cluster structure of $^{16}$O from first-principles calculations. In this letter, we present an \textit{ab initio} lattice 
calculation of the low-lying even-parity states of $^{16}$O using the framework of Nuclear Lattice Effective Field Theory (NLEFT), which combines
chiral nuclear EFT with lattice Monte Carlo simulations. From these considerations, we will provide evidence that the nucleons in the ground state of 
$^{16}$O are arranged in a tetrahedral configuration of alpha clusters. For the first excited $0^+$ state, we find a predominantly square configuration
of alpha clusters, the rotational excitations of which include the first $2^+$ state.

In chiral nuclear EFT, the interactions among nucleons are organized according to their importance based on a systematic expansion in powers of 
$Q/\Lambda$, where the ``hard scale'' $\Lambda \simeq 1$~GeV. The ``soft scale'' $Q$ is associated with nucleon three-momenta and 
the pion mass $m_\pi^{}$. 
The dominant contributions to the nuclear Hamiltonian appear at $\mathcal{O}(Q/\Lambda)^{0}$ or leading order (LO), while the 
next-to-leading order (NLO) terms are of $\mathcal{O}(Q/\Lambda)^{2}$ and involve the two-nucleon force (2NF) only.
In the results for $^{16}$O presented here, all relevant contributions to the nuclear Hamiltonian are taken into account up to
$\mathcal{O}(Q/\Lambda)^{3}$, or next-to-next-to-leading order (NNLO). In particular, this includes the three-nucleon force (3NF) which first appears at NNLO. The electromagnetic force, which is an important ingredient in nuclear binding, is
also included  consistently and systematically (for details, see Ref.~\cite{Epelbaum:2010xt}).
For recent reviews of chiral nuclear EFT, see Refs.~\cite{Epelbaum:2008ga,Machleidt:2011zz}. 

Our NLEFT calculations of $^{16}$O employ the same lattice action and algorithms previously used to study $^{12}$C with emphasis on 
the structure and quark mass dependence of the Hoyle state~\cite{Epelbaum:2011md,
Epelbaum:2012qn,Epelbaum:2012iu}, and for nuclei up to $A=28$~\cite{Lahde:2013uqa}. Our calculations use a periodic
cubic lattice with a spatial lattice spacing of $a=1.97$~fm and a length of $L=11.82$~fm. Euclidean time propagation is then used to project 
onto low-energy states of the $^{16}$O system. For any initial $A$-nucleon trial state $\Psi$, the projection amplitude is defined as the expectation 
value $\left\langle \exp(-Ht)\right\rangle_{\Psi}$, where $H$ denotes the Hamiltonian. For large Euclidean time $t$,
the exponential operator projects out the low-lying states, the energies of which are determined from the exponential decay of the corresponding 
projection amplitudes. In the Euclidean time direction, we have a temporal lattice spacing of $a_{t}^{}=1.32~$fm. The number of Euclidean time
steps $N_{t}^{}$ is varied in order to reach the limit $N_{t} \to \infty$ by extrapolation. Recent reviews of methods relevant to our NLEFT
calculations can be found in Refs.~\cite{Lee:2008fa,Drut:2012a}.

\begin{figure}[t]
\centering
\subfigure[\ Initial state ``A'', 8~equivalent~orientations.]
{\includegraphics[width=0.42\columnwidth]{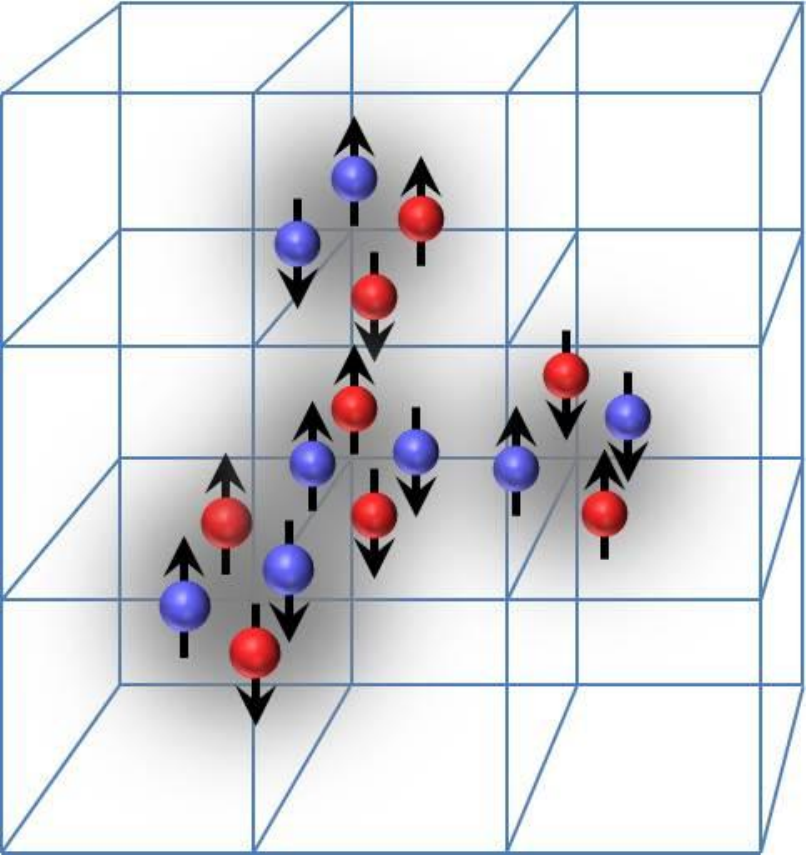} \label{fig:Tetra}}
\hspace{.5cm}
\subfigure[\ Initial states ``B'' and ``C'', 3~equivalent~orientations.]
{\includegraphics[width=0.42\columnwidth]{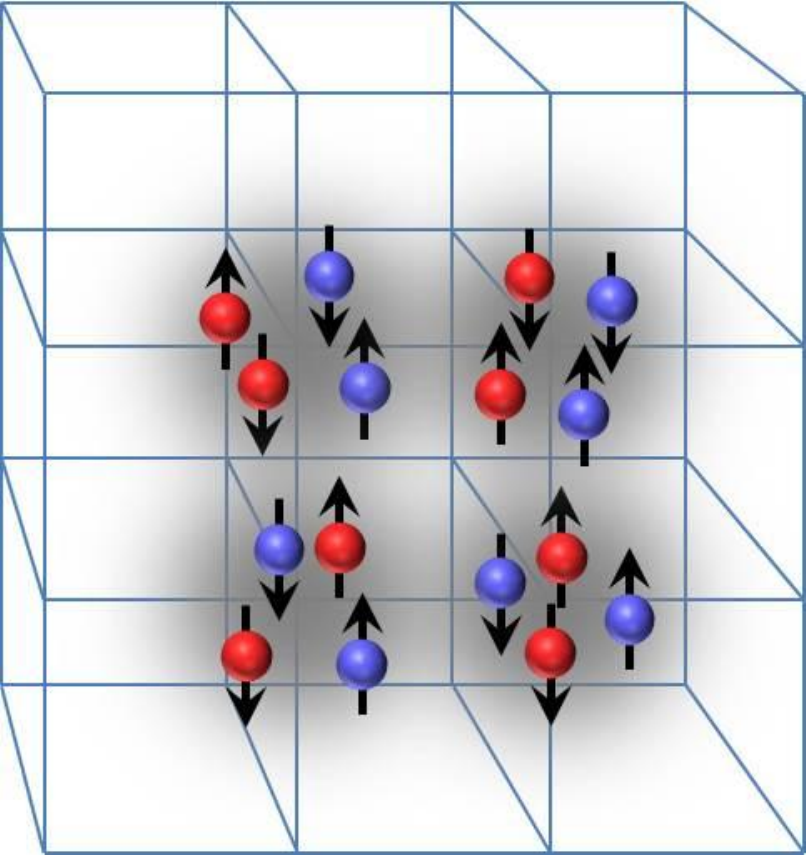} \label{fig:Square}}
\caption{Schematic illustration of the alpha cluster initial states with tetrahedral and square configurations.
Initial state ``C'' has the same geometry as ``B'' but with a larger radius for each of the four alpha clusters.
}
\end{figure}

The energy of the ground state of $^{16}$O was already calculated using NLEFT in Ref.~\cite{Lahde:2013uqa}, where multiple initial trial states
were employed in order to increase the accuracy of the $N_{t} \to \infty$ extrapolation. However, the structure of the ground state of $^{16}$O was
not explored. In order to gain insight into the structure of the lowest states in the spectrum of $^{16}$O, we investigate the Euclidean time evolution 
of specific initial trial states formed out of alpha clusters. For details on the implementation of such states on the lattice, see Ref.~\cite{Epelbaum:2012qn}. 
Our alpha cluster trial states are illustrated in Figs.~\ref{fig:Tetra} and~\ref{fig:Square}. In particular, we introduce a tetrahedral configuration of alpha 
clusters which we refer to as initial state ``A'', and a set of square configurations of alpha clusters. From the latter, we distinguish between initial 
state ``B'' where the alpha clusters are relatively compact, and initial state ``C'' where the alpha clusters have a greater spatial extent. In each case the alpha clusters are overlapping Gaussian distributions with an initial radius of $2.0$~fm for ``B'' and $2.8$~fm for ``C''.

\begin{figure}[t]
\includegraphics[width=\columnwidth]{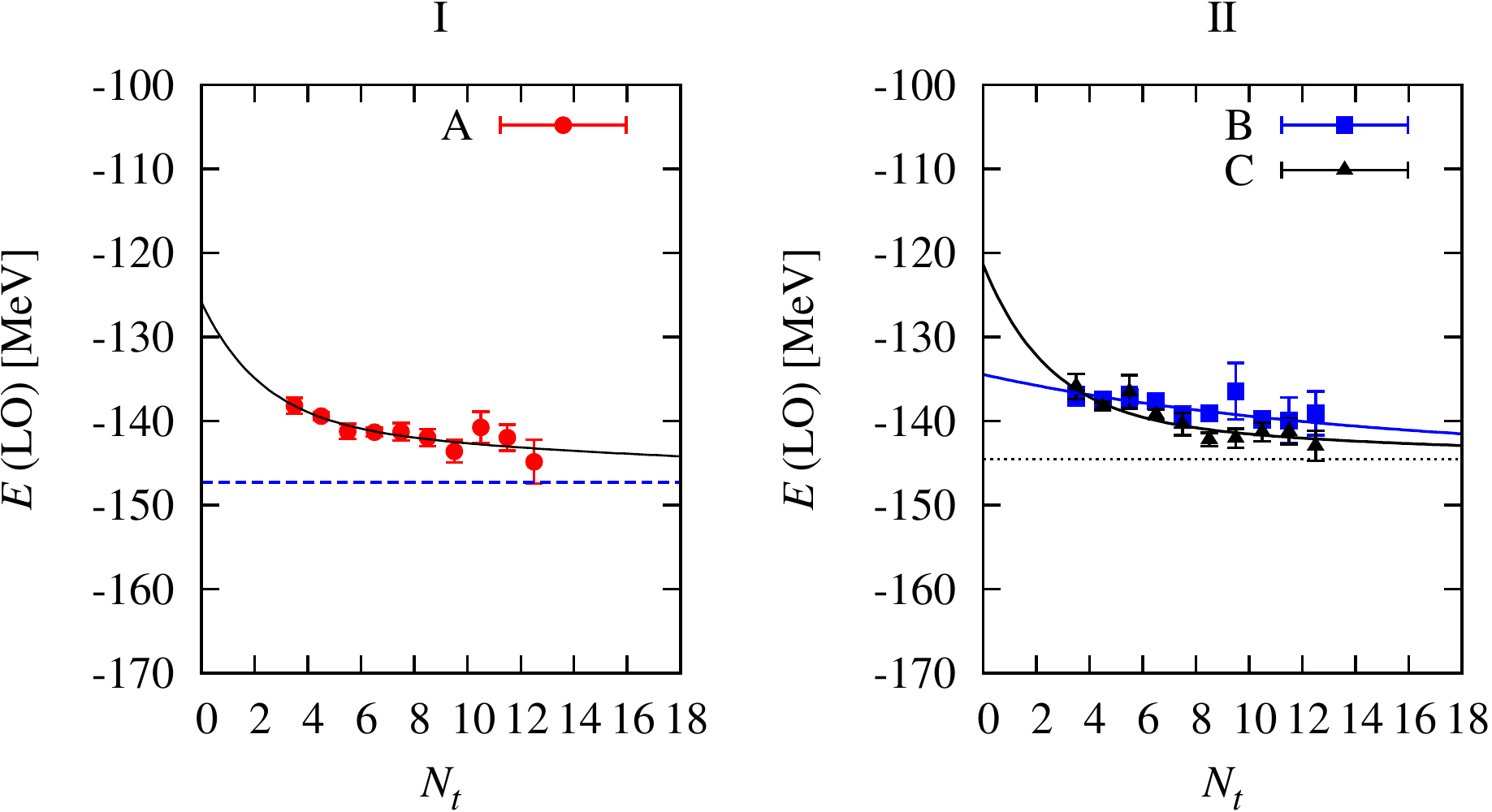}
\caption{NLEFT results for the LO energy as a function of Euclidean projection time. Panel~I shows the approach to the $0^+_1$ ground state
of $^{16}$O from initial state ``A'', and the dashed line shows the extrapolated value from Ref.~\cite{Lahde:2013uqa}. 
Panel~II shows the approach to the excited $0^+_2$ state from initial states ``B'' and ``C'', and the dotted line indicates the result of the
extrapolation $N_t^{} \to \infty$. These extrapolations are correlated with those for the higher-order corrections shown in 
Fig.~\ref{fig:NLONNLO}.
\label{fig:LO}}
\end{figure}

\begin{figure}[t]
\includegraphics[width=\columnwidth]{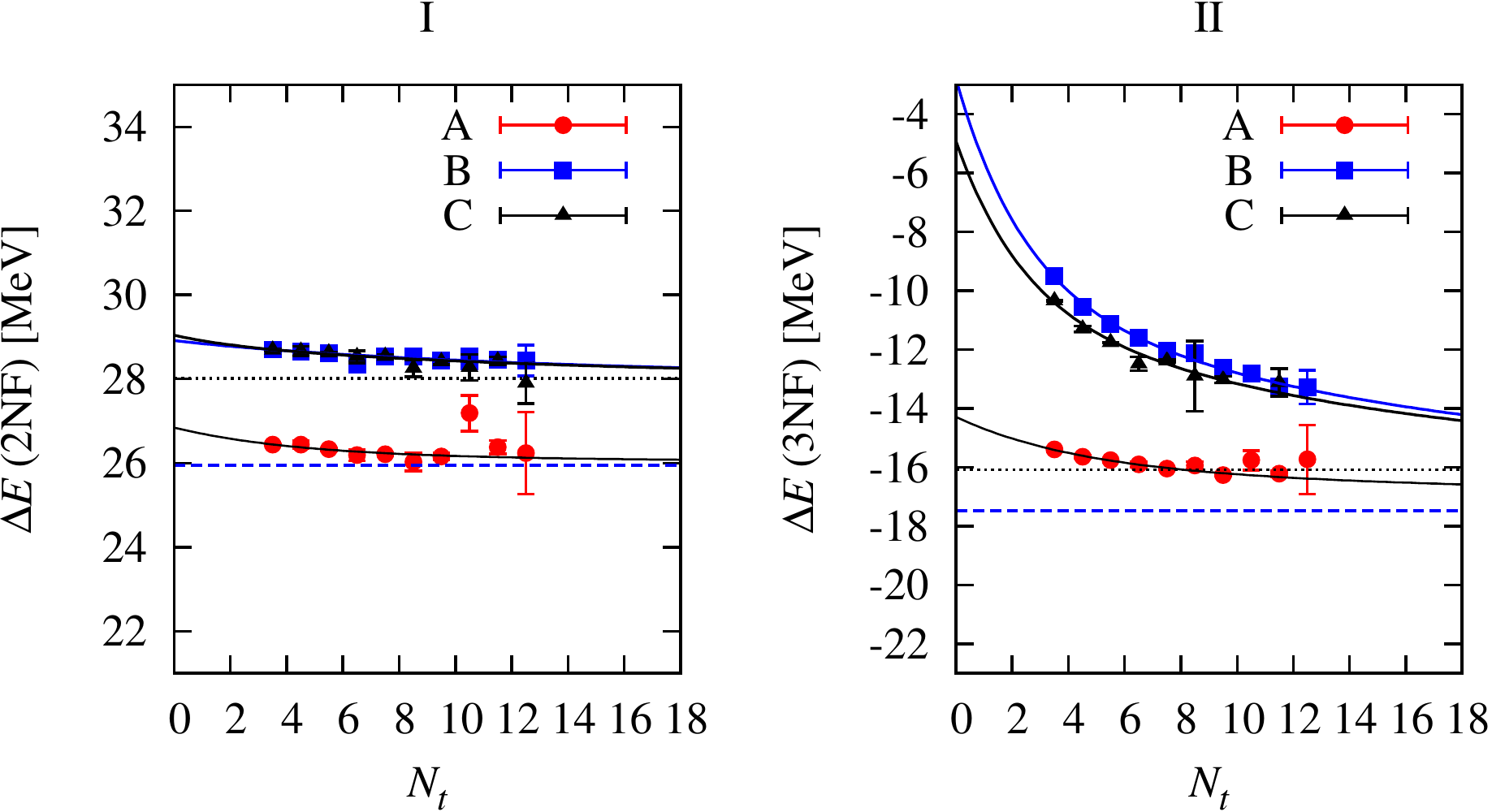}
\caption{NLEFT results for the higher-order corrections as a function of Euclidean projection time. The left panel gives the total contribution from
the 2NF up to NNLO, including electromagnetic and strong isospin breaking. The right panel shows the contribution from the 3NF
at NNLO (see Ref.~\cite{Lahde:2013uqa}).
Dashed lines indicate the extrapolated values for initial state ``A'', and the dotted lines indicate those for initial states ``B'' and ``C''. 
These extrapolations are correlated with those for the LO energies shown in Fig.~\ref{fig:LO}.
\label{fig:NLONNLO}}
\end{figure}

Our NLEFT results at LO are shown in Fig.~\ref{fig:LO}, where we plot the LO energy as a function of Euclidean projection time. The maximum extent
in $N_t^{}$ which can be explored without resorting to an extrapolation is limited by sign oscillations. The solid lines show exponential fits used for
the $N_t^{} \to \infty$ extrapolation (see Ref.~\cite{Lahde:2013uqa} for more details about this procedure). In Panel~I of Fig.~\ref{fig:LO}, we show our
NLEFT results obtained by starting the Euclidean time projection from a tetrahedral configuration of alpha clusters corresponding to initial state ``A''. The 
dashed horizontal line in Panel I of Fig.~\ref{fig:LO} shows the LO energy for the $0_1^+$ ground state of $^{16}$O found in Ref.~\cite{Lahde:2013uqa}, 
and the extrapolated energy for initial state ``A'' is completely consistent with the value $-147.3(5)~$MeV reported in Ref.~\cite{Lahde:2013uqa}.
We also find excellent agreement between the results based on initial state ``A'' and those reported in Ref.~\cite{Lahde:2013uqa} for the NLO and 
NNLO corrections to the ground state, shown in Fig.~\ref{fig:NLONNLO}.
   We find evidence for a $3^-$ rotational excitation of this tetrahedral configuration.  However, these results will be presented in a future publication on the odd-parity excitations of $^{16}$O.

In Panel~II of Fig.~\ref{fig:LO}, we present our NLEFT results for the LO energy based on Euclidean time projection from initial states ``B'' and ``C''. 
As will be shown below, these correspond to the excited $0_2^+$ state of $^{16}$O. The extrapolated LO energies for ``B'' and ``C'' 
give a common value of $-145(2)$~MeV, which is just slightly above the energy of the ground state. While there is some overlap between 
initial states ``B'' and ``C'' and the ground state, it is an order of magnitude smaller than for the $0_2^+$.  Therefore, we find a large window in $N_t^{}$ where the signal for the
$0_2^+$ state can be extracted without a full coupled-channel analysis.

\begin{table}[b]
\caption{NLEFT results and experimental (Exp) values for the lowest even-parity states of $^{16}$O (in MeV). 
The errors are one-standard-deviation estimates which include both statistical Monte Carlo errors and 
uncertainties due to the extrapolation $N_t^{} \to \infty$. The notation is identical to that of Ref.~\cite{Lahde:2013uqa}.
\label{tab_en}}
\begin{center}
\begin{tabular}{| c | r | r r r | r |}
\hline 
$J_n^p$ & \multicolumn{1}{c |}{LO (2N)} & \multicolumn{1}{c}{NNLO (2N)} 
& \multicolumn{1}{c}{+3N} & \multicolumn{1}{c |}{+4N$_\mathrm{eff}$} &
\multicolumn{1}{c |}{Exp} \\ \hline\hline
$0^+_1$ & $-147.3(5)$ & $-121.4(5)$ & $-138.8(5)$ & $-131.3(5)$ & $-127.62$ \\
$0^+_2$ & $-145(2)$ & $-116(2)$ & $-136(2)$ & $-123(2)$ & $-121.57$ \\
$2^+_1$ & $-145(2)$ & $-116(2)$ & $-136(2)$ & $-123(2)$ & $-120.70$ \\
\hline
\end{tabular}
\end{center}
\end{table}

\begin{figure}[t]
\includegraphics[width=\columnwidth]{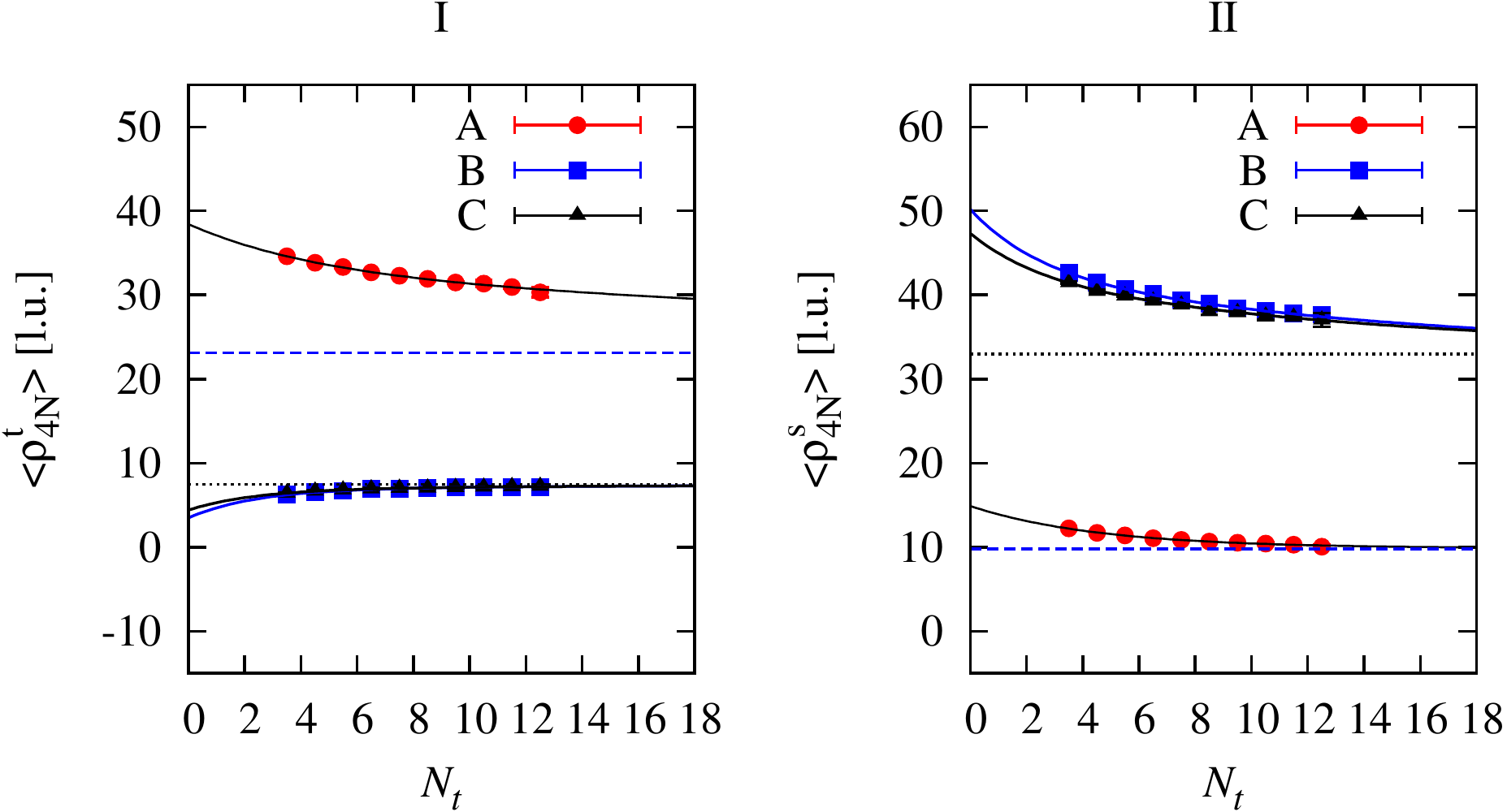}
\caption{NLEFT results for the 4N density operator expectation values $\langle\rho_\mathrm{4N}^t\rangle$ and $\langle\rho_\mathrm{4N}^s\rangle$
as a function of Euclidean projection time, for each of the trial wave functions employed. The dashed lines show the extrapolated values 
(from Ref.~\cite{Lahde:2013uqa}) equivalent to those of initial state ``A'', while the dotted lines show those corresponding to initial states ``B'' and ``C''.
Note the clear separation into the $0^+_1$ (initial state ``A'') and $0^+_2$ (initial states ``B'' and ``C'') states.
\label{fig:4N}}
\end{figure}

We are now in a position to verify that the ground state of $^{16}$O maintains the tetrahedral arrangement of alpha clusters characteristic of
initial state ``A'', and that the excited $0^+_2$ state maintains the square arrangement of alpha clusters in initial states ``B'' and ``C''.  In order to do this, 
we measure the expectation value of four-nucleon (4N) density operators, where each of the four nucleons are located on adjacent lattice sites, thus forming either 
a tetrahedron or a square. In Panel~I of Fig.~\ref{fig:4N}, we show the expectation value (in dimensionless lattice units, or l.u.)~of the tetrahedral density 
operator. The dashed horizontal line indicates the result
$\langle\rho_\mathrm{4N}^t\rangle \simeq 23.1(5)~$l.u.~from
the previous NLEFT calculations of the ground state of $^{16}$O in Ref.~\cite{Lahde:2013uqa}.  For initial state ``A'', $\langle\rho_\mathrm{4N}^t\rangle(t)$
for small $N_t^{}$ is somewhat larger than this value.  It, however, agrees perfectly with the quoted result in the limit $N_t^{} \to \infty$. We thus conclude that
a significant tetrahedral correlation of alpha clusters exists in the ground state of $^{16}$O. In contrast, $\langle\rho_\mathrm{4N}^t\rangle$ remains
roughly a factor of $\simeq 3$ smaller in the limit $N_t^{} \to \infty$ for initial states ``B'' and ``C''. Hence, it becomes clear that these trial wave functions
converge to a state distinct from the ground state under Euclidean time projection, which we identify as the excited $0^+_2$ state. Conversely, from
Panel~II of Fig.~\ref{fig:4N}, we find that the expectation value $\langle\rho_\mathrm{4N}^s\rangle$ of the square density operator is $\simeq 3$ times
larger for the $0^+_2$ state than for the ground state. Based on these results, we conclude that significant square-like correlations of alpha
clusters exist in the $0^+_2$ state of $^{16}$O.

In Table~\ref{tab_en}, we summarize our NLEFT results for the low-energy even-parity spectrum of $^{16}$O. The column labeled  
``LO (2N)'' refers to the LO energies, which depend on the 2NF only. We note that the LO results include some
higher order contributions due to the smearing of the 4N operators \cite{Epelbaum:2009zsa}. Similarly, ``NNLO (2N)'' takes into account all 2NF contributions
to the Hamiltonian up to NNLO in the chiral expansion, and ``+3N'' gives the result when the 3NF at NNLO
is accounted for.  For the employed improved lattice 
 implementation of the 3NF see Ref.~{\cite{Lahde:2013uqa}}. 
The column labeled ``+$\text{4N}_{\text{eff}}$'' gives our final result, after taking into account the effective 4N nearest-neighbor interaction
introduced in Ref.~\cite{Lahde:2013uqa}, which was tuned
to the empirical binding energy of $^{24}$Mg. It should be noted that all contributions at NLO and NNLO are treated in perturbation theory.

In addition to the $0^+_1$ and $0^+_2$ states, we also show in Table~\ref{tab_en} the energy of the $2^+_1$ state in the $E$ representation 
of the cubic symmetry group. This state is a rotational excitation of the $0^+_2$ state.  We find that the $E$ representation of the $2^+_1$ state 
is nearly degenerate with the $0^+_2$ state.  Overall, the empirical low-energy spectrum of $^{16}$O is reproduced relatively well. The NNLO results
with the 3NF included show an overbinding of $\simeq 10\%$, and a somewhat too small excitation energy for the $0^+_2$ and $2^+_1$ states.
While these problems are remedied by the effective 4N interaction introduced in Ref.~\cite{Lahde:2013uqa}, in future studies the necessary corrections
should be provided by a combination of the hitherto missing next-to-next-to-next-to-leading order (N3LO) terms and a reduced lattice spacing.

In Table~\ref{tab_trans}, we first show the NLEFT results for the
charge radii of the low-energy even-parity states of $^{16}$O at LO. We find that the LO result for the ground state  
is $\simeq 20\%$ smaller than the empirical value. This result is consistent with the observed $\simeq 20$~MeV overbinding at LO. The charge radii
are expected to increase significantly as the energy is pushed closer to the $^{12}\text{C}+\alpha$ threshold, especially for the $0^+_2$ and $2^+_1$ 
states which are close to that threshold. Unfortunately, calculations of the higher-order corrections to the charge radii and other electromagnetic 
observables are computationally significantly more expensive than the LO calculations, and therefore go beyond the scope of the current analysis. 
We are working on the inclusion of these higher-order corrections in a future publication. 


We subsequently give in Table~\ref{tab_trans} the NLEFT results at LO
for the electric quadrupole moment of the $2^+_1$ state, the electric quadrupole ($E2$) 
transition probabilities, and the electric monopole ($E0$) matrix element.  Since the LO
charge radius $r_{\text{LO}}^{}$ of the ground state is smaller than the 
empirical value $r_{\text{exp}}^{}$, a systematic deviation appears, which arises from the overall size of the second moment of the charge distribution. 
To compensate for this overall scaling mismatch, we have also calculated ``rescaled'' quantities multiplied by powers of the ratio 
$r_{\text{exp}}^{}/r_{\text{LO}}^{}$, according to the length dimension of each observable.

\begin{table}[t]
\caption{NLEFT results for the charge radius $r$, the quadrupole moment $Q$, and the electromagnetic transition amplitudes for $E2$ and 
$E0$ transitions, as defined in Ref.~\cite{Mottelson_book}.
We compare with empirical (Exp) values where these are known. For the quadrupole moment and the transition amplitudes, we also show
``rescaled'' LO results, which correct for the deviation from the empirical value of the charge radius at LO (see main text). 
The uncertainties are one-standard-deviation estimates which include the statistical Monte Carlo error as well as the errors due 
to the $N_t^{} \to \infty$ extrapolation.
\label{tab_trans}}
\begin{center}
\begin{tabular}{| c | c c | c |}
\hline  
& LO & rescaled & Exp \\ \hline\hline
$r(0^+_1)$ [fm] & 2.3(1) & --- & 2.710(15) \cite{Kim:1978st} \\
$r(0^+_2)$ [fm] & 2.3(1) & --- & --- \\
$r(2^+_1)$ [fm] & 2.3(1) & --- & --- \\
$Q(2^+_1)$ [$e$ fm$^2$] & 10(2) & 15(3) & --- \\
$B(E2, 2^+_1 \to 0^+_2)$ [$e^2$fm$^4$] & 22(4) & 46(8) & 65(7) \cite{Ajzenberg:1971}\\
$B(E2, 2^+_1 \to 0^+_1)$ [$e^2$fm$^4$] & 3.0(7) & 6.2(1.6) & 7.4(2) \cite{Moreh:1985zz}\\
$M(E0, 0^+_2 \to 0^+_1)$ [$e$ fm$^2$] & 2.1(7) & 3.0(1.4) & 3.6(2) \cite{Miska:1975} \\
\hline 
\end{tabular}
\end{center}
\end{table}

With the scaling factor included, we find that the NLEFT predictions for the $E2$ and 
$E0$ transitions are in good agreement with the experimental values. 
In particular, NLEFT is able to explain the empirical value of $B(E2,2^+_1 \to 0^+_2)$, which is $\simeq 30$ times larger than the Weisskopf 
single-particle shell model estimate. This provides further confirmation of the interpretation of the $2^+_1$ state as a rotational excitation of the 
$0^+_2$ state. Finally, we provide a prediction for the quadrupole moment of the $2^+_1$ state. We note that the NLEFT calculation of the 
electromagnetic transitions requires a full coupled-channel analysis.  For such calculations, we use initial states that consist of a compact triangle of 
alpha clusters and a fourth alpha cluster, located either in the plane of the triangle (square-like) or out of the plane of the triangle (tetrahedral).

We should mention that all of the low-energy states of $^{16}$O discussed in this letter can also be obtained by Euclidean time projection  acting upon initial states with no alpha clustering at all.  We can measure the degree of alpha cluster formation by calculating the local four-nucleon density as a function of projection time.  For non-alpha-cluster initial states, the local four-nucleon density starts very low and then increases substantially with projection time.  For alpha-cluster initial states, however, the local four-nucleon density starts much higher and then remains elevated as a function of projection time.  This gives us confidence that the observed formation of alpha clusters in our lattice simulations are not produced a particular choice of initial states but rather the result of strong four-nucleon correlations in the $^{16}$O  system.

In summary, we have presented $\textit{ab initio}$ results for the low-energy even-parity states of $^{16}$O using NLEFT, that are in good agreement
with available empirical data for the energy spectrum and electromagnetic properties. We have also made advances in the understanding of the
structure of $^{16}$O. In particular, we have presented an \textit{ab initio} confirmation of the underlying structures of the ground state and the first 
excited state. For the ground state, we find that the nucleons are dominantly arranged in a tetrahedral configuration of alpha clusters. For the first 
excited state, the predominant structure is a square-like configuration of alpha clusters, with rotational excitations that include the first spin-2 state.  
Much remains to be studied in the $^{16}$O system, such as the computation of the odd-parity spectrum and the inclusion of corrections 
beyond LO for the electromagnetic observables. We also plan to decrease the lattice spacing and to include the N3LO corrections. This should
enable us to describe the spectrum of $^{16}$O to better accuracy without an effective 4N interaction.

\medskip

\begin{acknowledgments}
We acknowledge partial financial support from the Deutsche Forschungsgemeinschaft  and NSFC (Sino-German CRC 110), the Helmholtz Association 
(Contract No.\ VH-VI-417), BMBF (Grant No.\ 05P12PDFTE), the NSF\  (PHY-1307453), and the DOE (DE-FG02-03ER41260). Further support
was provided by the EU HadronPhysics3 project and the ERC Project No.\ 259218 NUCLEAREFT. The computational resources 
were provided by the J\"{u}lich Supercomputing Centre at  Forschungszentrum J\"{u}lich and by RWTH Aachen.
\end{acknowledgments}



\begin{thebibliography}{99}

\bibitem{Hagen:2010gd}
G.~Hagen, T.~Papenbrock, D.~J.~Dean, and M.~Hjorth-Jensen,
Phys.\ Rev.\ C {\bf 82}, 034330 (2010).

\bibitem{Hergert:2013uja}
H.~Hergert, S.~Binder, A.~Calci, J.~Langhammer, and R.~Roth,
Phys.\ Rev.\ Lett.\ {\bf 110}, 242501 (2013).

\bibitem{Roth:2011ar} 
  R.~Roth,  J.~Langhammer, A.~Calci, S.~Binder, P.~Navratil,
  Phys.\ Rev.\ Lett.\ {\bf 107}, 072501 (2011).

\bibitem{Nagai:1962} 
  H.~Nagai,
  Prog.\ Theor.\ Phys.\ {\bf 27}, 619 (1962).

\bibitem{Wheeler:1937zza} 
  J.~A.~Wheeler,
  Phys.\ Rev.\ {\bf 52}, 1083 (1937).

\bibitem{Dennison:1954zz}
  D.~M.~Dennison,
  Phys.\ Rev.\ {\bf 96}, 378 (1954).
  
\bibitem{Iachello}
  H.~Feshbach and F.~Iachello,
  Phys.\ Lett.\ B {\bf 45}, 7 (1973).
  
\bibitem{Robson:1979zz}
  D.~Robson,
  Phys.\ Rev.\ Lett.\ {\bf 42}, 876 (1979).

\bibitem{Bauhoff:1984zza}
  W.~Bauhoff, H.~Schultheis, and R.~Schultheis,
  Phys.\ Rev.\ C {\bf 29}, 1046 (1984).

\bibitem{Filikhin:1999}
  I.~N.~Filikhin and S.~L.~Yakovlev,
  Phys.\ Atom.\ Nucl.\ {\bf 64}, 409 (2000).

\bibitem{Tohsaki:2001an}
  A.~Tohsaki, H.~Horiuchi, P.~Schuck, and G.~R\"opke,
  Phys.\ Rev.\ Lett.\ {\bf 87}, 192501 (2001).

\bibitem{Bijker}
  R.~Bijker, 
  AIP Conf.\ Proc.\ {\bf 1323}, 28 (2010);
  {\it ibid.}, J.\ Phys.:\ Conf.\ Ser.\ {\bf 380}, 012003 (2012).
  
\bibitem{Freer:2005ia}
  M.~Freer (CHARISSA Collaboration),
  J.\ Phys.\ G {\bf 31}, S1795 (2005).

\bibitem{Buck:1975zz} 
  B.~Buck, C.~B.~Dover and J.~P.~Vary,
  Phys.\ Rev.\ C {\bf 11}, 1803 (1975).
  
\bibitem{Epelbaum:2008ga}
  E.~Epelbaum, H.-W.~Hammer, and U.-G.~Mei{\ss}ner,
  Rev.\ Mod.\ Phys.\ {\bf 81}, 1773 (2009).
  
\bibitem{Machleidt:2011zz}
  R.~Machleidt and D.~R.~Entem,
  Phys.\ Rept.\ {\bf 503}, 1 (2011).

\bibitem{Epelbaum:2011md} 
  E.~Epelbaum, H.~Krebs, D.~Lee, and U.-G.~Mei{\ss}ner,
  Phys.\ Rev.\ Lett.\  {\bf 106}, 192501 (2011).

\bibitem{Epelbaum:2012qn} 
  E.~Epelbaum, H.~Krebs, T.~A.~L\"ahde, D.~Lee, and U.-G.~Mei{\ss}ner,
  Phys.\ Rev.\ Lett.\ {\bf 109}, 252501 (2012).

\bibitem{Epelbaum:2012iu} 
  E.~Epelbaum, H.~Krebs, T.~A.~L\"ahde, D.~Lee, and U.-G.~Mei{\ss}ner,
  Phys.\ Rev.\ Lett.\ {\bf 110}, 112502 (2013);
  {\it ibid.}, Eur.\ Phys.\ J.\ A {\bf 49}, 82 (2013).

\bibitem{Lahde:2013uqa}
  T.~A.~L\"ahde, E.~Epelbaum, H.~Krebs, D.~Lee, U.-G.~Mei{\ss}ner, and G.~Rupak,
  arXiv:1311.0477 [nucl-th].
 
\bibitem{Epelbaum:2010xt} 
  E.~Epelbaum, H.~Krebs, D.~Lee and U.-G.~Mei{\ss}ner,
  Eur.\ Phys.\ J.\ A {\bf 45}, 335 (2010)
  [arXiv:1003.5697 [nucl-th]].
  
\bibitem{Lee:2008fa}
  D.~Lee,
  Prog.\ Part.\ Nucl.\ Phys. {\bf 63}, 117 (2009).

\bibitem{Drut:2012a}
  J.~E.~Drut and A.~N.~Nicholson,
  J.\ Phys.\ G {\bf 40}, 043101 (2013).

\bibitem{Epelbaum:2009zsa} 
  E.~Epelbaum, H.~Krebs, D.~Lee and U.-G.~Mei{\ss}ner,
  Eur.\ Phys.\ J.\ A {\bf 41}, 125 (2009).

\bibitem{Mottelson_book}
  A.~Bohr and B.~R.~Mottelson,
  {\it Nuclear Structure. Single-Particle Motion}
  (W.~A.~Benjamin, New York, 1969), Vol.~I.

\bibitem{Kim:1978st}
  J.~C.~Kim {\it et al.},
  Nucl.\ Phys.\ {\bf A297}, 301 (1978).

\bibitem{Ajzenberg:1971}
  F.~Ajzenberg-Selove,
  Nucl.\ Phys.\ {\bf B166}, 1 (1971).
  
\bibitem{Moreh:1985zz}
  R.~Moreh, W.~C.~Sellyey, D.~Sutton, and R.~Vodhanel,
  Phys.\ Rev.\ C {\bf 31}, 2314 (1985).

\bibitem{Miska:1975}
  H.~Miska {\it et al.},
  Phys.\ Lett.\ B {\bf 58}, 155 (1975).

\end{thebibliography}

\end{document}